\documentclass[longauth]{aa}

\usepackage{amsmath}
\usepackage{natbib}
\usepackage{graphicx}
\usepackage[varg]{txfonts}
\usepackage{color}
\usepackage{url}
\usepackage{hyperref}
\usepackage{amssymb}

\usepackage[utf8]{inputenc}
\usepackage[T1]{fontenc}
\usepackage{textcomp}
\DeclareUnicodeCharacter{0327}{\textcommabelow}

\setcitestyle{citesep={,}}

\def\aj{AJ}

\def\apj{ApJ}
\def\apjl{ApJL}
\def\apjs{ApJS}
\def\apss{Ap\&SS}
\def\aap{A\&A}

\def\mnras{MNRAS}

\def\prl{Phys.~Rev.~Lett.}

\def\pasj{PASJ}
\def\prd{Phys.~Rev.~D.}
\def\nat{Nature}

\def\physrep{Phys.~Rep.}

\usepackage{wasysym}

\begin{document}

%\linenumbers

\title{H.E.S.S. observations of composite Seyfert–starburst galaxies}

\authorrunning{H.E.S.S. Collaboration}
\titlerunning{Composite galaxies observed with H.E.S.S.}

\author{\fontsize{11.0}{13.0}\selectfont{H.E.S.S. Collaboration
\and A.~Acharyya\inst{\ref{I1}}  % \orcidlink{0000-0002-2028-9230}, HESS member
\and F.~Aharonian\inst{\ref{I2},\ref{I3}}  % \orcidlink{0000-0003-1157-3915}, HESS member
\and H.~Ashkar\inst{\ref{I4}}  % \orcidlink{0000-0002-2153-1818}, HESS member
\and M.~Backes\inst{\ref{I5},\ref{I6}}  % \orcidlink{0000-0002-9326-6400}, HESS member
\and V.~Barbosa~Martins\inst{\ref{I7}}  % \orcidlink{0000-0002-5085-8828}, HESS member
\and R.~Batzofin\inst{\ref{I8}}  % \orcidlink{0000-0002-5797-3386}, HESS member
\and Y.~Becherini\inst{\ref{I9},\ref{I10}}  % \orcidlink{0000-0002-2115-2930}, HESS member
\and D.~Berge\inst{\ref{I7},\ref{I11}}  % \orcidlink{0000-0002-2918-1824}, HESS member
\and M.~B\"ottcher\inst{\ref{I6}}  % \orcidlink{0000-0002-8434-5692}, HESS member
\and C.~Boisson\inst{\ref{I12}}  % \orcidlink{0000-0001-5893-1797}, HESS member
\and J.~Bolmont\inst{\ref{I13}}  % \orcidlink{0000-0003-4739-8389}, HESS member
\and J.~Borowska\inst{\ref{I11}}  % HESS member
\and F.~Brun\inst{\ref{I14}}  % \orcidlink{0000-0003-0770-9007}, HESS member
\and B.~Bruno\inst{\ref{I15}}  % HESS member
\and C.~Burger-Scheidlin\inst{\ref{I2}}  % \orcidlink{0000-0002-7239-2248}, HESS member
\and S.~Casanova\inst{\ref{I16}}  % \orcidlink{0000-0002-6144-9122}, HESS member
\and J.~Celic\inst{\ref{I15}}  % HESS member
\and M.~Cerruti\inst{\ref{I9}}  % \orcidlink{0000-0001-7891-699X}, HESS member
\and S.~Chandra\inst{\ref{I6}}  % HESS member
\and A.~Chen\inst{\ref{I17},}\protect\footnotemark[1]  % \orcidlink{0000-0001-6425-5692}, HESS member
\and M.~Chernyakova\inst{\ref{I2_1}, \ref{I2}}  % \orcidlink{0000-0002-9735-3608}, HESS member
\and J.~O.~Chibueze\inst{\ref{I6},\ref{I5}}  % \orcidlink{0000-0002-9875-7436}, HESS member
\and O.~Chibueze\inst{\ref{I6}}  % \orcidlink{0000-0001-8601-2675}, HESS member
\and S.~Colafrancesco\inst{\ref{I17},}\protect\footnotemark[2]
\and T.~Collins\inst{\ref{I8}}  % \orcidlink{0000-0001-5020-5387}, HESS member
\and B.~Cornejo\inst{\ref{I14}}  % \orcidlink{0009-0003-0039-0483}, HESS member
\and G.~Cotter\inst{\ref{I18}}  % \orcidlink{0000-0002-9975-1829}, HESS member
\and J.~Damascene~Mbarubucyeye\inst{\ref{I7}}  % \orcidlink{0000-0002-4991-6576}, HESS member
\and J.~de~Assis~Scarpin\inst{\ref{I4}}  % \orcidlink{0009-0004-4411-236X}, HESS member
\and M.~de~Bony~de~Lavergne\inst{\ref{I14},\ref{I19}}  % \orcidlink{0000-0002-4650-1666}, HESS member
\and M.~de~Naurois\inst{\ref{I4}}  % \orcidlink{0000-0002-7245-201X}, HESS member
\and E.~de~O\~na~Wilhelmi\inst{\ref{I7}}  % \orcidlink{0000-0002-5401-0744}, HESS member
\and A.~G.~Delgado~Giler\inst{\ref{I11}}  % \orcidlink{0000-0003-2190-9857}, HESS member
\and A.~Djannati-Ata\"i\inst{\ref{I9}}  % \orcidlink{0000-0002-4924-1708}, HESS member
\and J.~Djuvsland\inst{\ref{I3}}  % \orcidlink{0000-0002-6488-8219}, HESS member
\and A.~Dmytriiev\inst{\ref{I6}}  % \orcidlink{0000-0003-0102-5579}, HESS member
\and R.~Ebrahim\inst{\ref{I17}}
\and K.~Egg\inst{\ref{I15}}  % \orcidlink{0009-0002-4238-034X}, HESS member
\and C.~Esca\~{n}uela~Nieves\inst{\ref{I3}}  % \orcidlink{0000-0002-7297-8126}, HESS member
\and K.~Feijen\inst{\ref{I9}}  % \orcidlink{0000-0003-1476-3714}, HESS member
\and M.~D.~Filipovic\inst{\ref{I20}}  % \orcidlink{0000-0002-4990-9288}, HESS member
\and G.~Fontaine\inst{\ref{I4}}  % \orcidlink{0000-0002-6443-5025}, HESS member
\and S.~Funk\inst{\ref{I15}}  % \orcidlink{0000-0002-2012-0080}, HESS member
\and S.~Gabici\inst{\ref{I9}}  % HESS member
\and J.F.~Glicenstein\inst{\ref{I14}}  % \orcidlink{0000-0003-2581-1742}, HESS member
\and J.~Glombitza\inst{\ref{I15}}  % HESS member
\and P.~Goswami\inst{\ref{I21}}  % \orcidlink{0000-0001-5430-4374}, HESS member
\and M.-H.~Grondin\inst{\ref{I22}}  % \orcidlink{0000-0002-8383-251X}, HESS member
\and B.~Hess\inst{\ref{I23}}  % \orcidlink{0009-0004-9999-171X}, HESS member
\and J.A.~Hinton\inst{\ref{I3}}  % \orcidlink{0000-0002-1031-7760}, HESS member
\and T.~L.~Holch\inst{\ref{I7},}\protect\footnotemark[1]  % \orcidlink{0000-0001-5161-1168}, HESS member
\and M.~Holler\inst{\ref{I24}}  % \orcidlink{0000-0002-0107-8657}, HESS member
\and D.~Horns\inst{\ref{I25}}  % \orcidlink{0000-0003-1945-0119}, HESS member
\and M.~Jamrozy\inst{\ref{I26}}  % \orcidlink{0000-0002-0870-7778}, HESS member
\and F.~Jankowsky\inst{\ref{I21}}  % HESS member
\and I.~Jung-Richardt\inst{\ref{I15}}  % HESS member
\and E.~Kasai\inst{\ref{I5}}  % \orcidlink{0000-0001-9696-7221}, HESS member
\and K.~Kasprzak\inst{\ref{I26}}  % \orcidlink{0009-0006-8475-9074}, HESS member
\and K.~Katarzy{\'n}ski\inst{\ref{I27}}  % \orcidlink{0000-0002-8806-4863}, HESS member
\and D.~Kerszberg\inst{\ref{I13}}  % \orcidlink{0000-0002-5289-1509}, HESS member
\and B. Khélifi\inst{\ref{I9}}  % \orcidlink{0000-0001-6876-5577}, HESS member
\and N.~Komin\inst{\ref{I28},\ref{I17},}\protect\footnotemark[1]  % \orcidlink{0000-0003-3280-0582}, HESS member
\and K.~Kosack\inst{\ref{I14}}  % \orcidlink{0000-0001-8424-3621}, HESS member
\and D.~Kostunin\inst{\ref{I7}}  % \orcidlink{0000-0002-0487-0076}, HESS member
\and R.G.~Lang\inst{\ref{I15}}  % \orcidlink{0000-0003-0492-5628}, HESS member
\and S.~Lazarevi\'c\inst{\ref{I20}}  % \orcidlink{0000-0001-6109-8548}, HESS member
\and V.~Lefranc\inst{\ref{I14},}\protect\footnotemark[1]
\and J.-P.~Lenain\inst{\ref{I13}}  % \orcidlink{0000-0001-7284-9220}, HESS member
\and P.~Liniewicz\inst{\ref{I26}}  % \orcidlink{0009-0008-3575-3965}, HESS member
\and A.~Luashvili\inst{\ref{I6}}  % \orcidlink{0000-0003-4384-1638}, HESS member
\and J.~Mackey\inst{\ref{I2}}  % \orcidlink{0000-0002-5449-6131}, HESS member
\and D.~Malyshev\inst{\ref{I23}}  % \orcidlink{0000-0001-9689-2194}, HESS member
\and D.~Malyshev\inst{\ref{I15}}  % \orcidlink{0000-0002-9102-4854}, HESS member
\and V.~Marandon\inst{\ref{I14}}  % \orcidlink{0000-0001-9077-4058}, HESS member
\and P.~Marchegiani\inst{\ref{I17},\ref{l29}}
\and M.~Mayer\inst{\ref{I15}}  % HESS member
\and A.~Mehta\inst{\ref{I7}}  % HESS member
\and A.M.W.~Mitchell\inst{\ref{I15}}  % \orcidlink{0000-0003-3631-5648}, HESS member
\and R.~Moderski\inst{\ref{I30}}  % \orcidlink{0000-0002-8663-3882}, HESS member
\and M.O.~Moghadam\inst{\ref{I8}}  % \orcidlink{0009-0003-2479-1863}, HESS member
\and L.~Mohrmann\inst{\ref{I3}}  % \orcidlink{0000-0002-9667-8654}, HESS member
\and E.~Moulin\inst{\ref{I14}}  % \orcidlink{0000-0003-4007-0145}, HESS member
\and J.~Niemiec\inst{\ref{I16}}  % \orcidlink{0000-0001-6036-8569}, HESS member
\and P.~O'Brien\inst{\ref{I31}}  % \orcidlink{0000-0002-5128-1899}, HESS member
\and L.~Olivera-Nieto\inst{\ref{I3}}  % \orcidlink{0000-0002-9105-0518}, HESS member
\and S.~Panny\inst{\ref{I24}}  % \orcidlink{0000-0001-5770-3805}, HESS member
\and M.~Panter\inst{\ref{I3}}  % HESS member
\and R.D.~Parsons\inst{\ref{I11}}  % \orcidlink{0000-0003-3457-9308}, HESS member
\and U.~Pensec\inst{\ref{I13}}  % \orcidlink{0009-0009-2359-1775}, HESS member
\and D.~A.~Prokhorov\inst{\ref{WU},}\protect\footnotemark[1]
\and G.~P\"uhlhofer\inst{\ref{I23}}  % \orcidlink{0000-0003-4632-4644}, HESS member
\and M.~Punch\inst{\ref{I9}}  % \orcidlink{0000-0002-4710-2165}, HESS member
\and A.~Quirrenbach\inst{\ref{I21}}  % HESS member
\and M.~Regeard\inst{\ref{I9}}  % \orcidlink{0000-0002-3844-6003}, HESS member
\and A.~Reimer\inst{\ref{I24}}  % \orcidlink{0000-0001-8604-7077}, HESS member
\and O.~Reimer\inst{\ref{I24}}  % \orcidlink{0000-0001-6953-1385}, HESS member
\and H.~Ren\inst{\ref{I3}}  % \orcidlink{0000-0003-0221-2560}, HESS member
\and F.~Rieger\inst{\ref{I3}}  % HESS member
\and B.~Rudak\inst{\ref{I30}}  % \orcidlink{0000-0003-0452-3805}, HESS member
\and K.~Sabri\inst{\ref{I28}}  % HESS member
\and V.~Sahakian\inst{\ref{I32}}  % \orcidlink{0000-0003-1198-0043}, HESS member
\and H.~Salzmann\inst{\ref{I23}}  % \orcidlink{0009-0000-0295-8800}, HESS member
\and M.~Sasaki\inst{\ref{I15}}  % \orcidlink{0000-0001-5302-1866}, HESS member
\and F.~Sch\"ussler\inst{\ref{I14}}  % \orcidlink{0000-0003-1500-6571}, HESS member
\and J.N.S.~Shapopi\inst{\ref{I5}}  % \orcidlink{0000-0002-7130-9270}, HESS member
\and W.~Si~Said\inst{\ref{I4}}  % \orcidlink{0009-0007-6555-6893}, HESS member
\and S.~Spencer\inst{\ref{I15}}  % \orcidlink{0000-0001-5516-1205}, HESS member
\and {\L.}~Stawarz\inst{\ref{I26}}  % \orcidlink{0000-0002-7263-7540}, HESS member
\and S.~Steinmassl\inst{\ref{I3}}  % \orcidlink{0000-0002-2865-8563}, HESS member
\and T.~Tanaka\inst{\ref{I33}}  % \orcidlink{0000-0002-4383-0368}, HESS member
\and A.M.~Taylor\inst{\ref{I7}}  % \orcidlink{0000-0001-9473-4758}, HESS member
\and R.~Terrier\inst{\ref{I9}}  % \orcidlink{0000-0002-8219-4667}, HESS member
\and M.~Tsirou\inst{\ref{I7}}  % \orcidlink{0000-0003-3417-1425}, HESS member
\and T.~Unbehaun\inst{\ref{I15}}  % \orcidlink{0000-0002-7378-4024}, HESS member
\and C.~van~Eldik\inst{\ref{I15}}  % \orcidlink{0000-0001-9669-645X}, HESS member
\and M.~Vecchi\inst{\ref{I34}}  % \orcidlink{0000-0002-5338-6029}, HESS member
\and C.~Venter\inst{\ref{I6}}  % \orcidlink{0000-0002-2666-4812}, HESS member
\and J.~Vink\inst{\ref{I35}}  % \orcidlink{0000-0002-4708-4219}, HESS member
\and T.~Wach\inst{\ref{I15}}  % HESS member
\and A.~Wierzcholska\inst{\ref{I16},\ref{I21}}  % \orcidlink{0000-0003-4472-7204}, HESS member
\and A.~Zech\inst{\ref{I12}}  % HESS member
\and W.~Zhong\inst{\ref{I7}}  % \orcidlink{0000-0003-3717-2861}, HESS member
}
}
\institute{
University of Southern Denmark, Campusvej 55, DK-5230 Odense M, Denmark \label{I1}
\and Astronomy \& Astrophysics Section, School of Cosmic Physics, Dublin Institute for Advanced Studies, DIAS Dunsink Observatory, Dublin D15 XR2R, Ireland \label{I2}
\and Max-Planck-Institut für Kernphysik, P.O. Box 103980, D 69029 Heidelberg, Germany \label{I3}
\and Laboratoire Leprince-Ringuet, École Polytechnique, CNRS, Institut Polytechnique de Paris, F-91128 Palaiseau, France \label{I4}
\and University of Namibia, Department of Physics, Private Bag 13301, Windhoek 10005, Namibia \label{I5}
\and Centre for Space Research, North-West University, Potchefstroom 2520, South Africa \label{I6}
\and Deutsches Elektronen-Synchrotron DESY, Platanenallee 6, 15738 Zeuthen, Germany \label{I7}
\and Institut für Physik und Astronomie, Universität Potsdam, Karl-Liebknecht-Strasse 24/25, D 14476 Potsdam, Germany \label{I8}
\and Université Paris Cité, CNRS, Astroparticule et Cosmologie, F-75013 Paris, France \label{I9}
\and Department of Physics and Electrical Engineering, Linnaeus University, 351 95 Växjö, Sweden  \label{I10}
\and Institut für Physik, Humboldt-Universität zu Berlin, Newtonstr. 15, D 12489 Berlin, Germany \label{I11}
\and LUX, Observatoire de Paris, Université PSL, CNRS, Sorbonne Université, 5 Pl. Jules Janssen, 92190 Meudon, France \label{I12}
\and Sorbonne Université, CNRS/IN2P3, Laboratoire de Physique Nucléaire, et de Hautes Energies, LPNHE, 4 place Jussieu, 75005 Paris, France \label{I13}
\and IRFU, CEA, Université Paris-Saclay, F-91191 Gif-sur-Yvette, France \label{I14}
\and Friedrich-Alexander-Universität Erlangen-Nürnberg, Erlangen Centre for Astroparticle Physics,  Nikolaus-Fiebiger-Str. 2, 91058 Erlangen, Germany \label{I15}
\and Instytut Fizyki Ja̧drowej PAN, ul. Radzikowskiego 152, ul. Radzikowskiego 152, 31-342 Kraków, Poland \label{I16}
\and School of Physics, University of the Witwatersrand, 1 Jan Smuts Avenue, Braamfontein, Johannesburg, 2050, South Africa \label{I17}
\and School of Physical Sciences and Centre for Astrophysics \& Relativity, Dublin City University, Glasnevin, D09 W6Y4, Ireland \label{I2_1}
\and University of Oxford, Department of Physics, Denys Wilkinson Building, Keble Road, Oxford OX1 3RH, UK, UK \label{I18}
\and Aix Marseille Université, CNRS/IN2P3, CPPM, Marseille, France \label{I19}
\and School of Science, Western Sydney University, Locked Bag 1797, Penrith South DC, NSW 2751, Australia \label{I20}
\and Landessternwarte, Universität Heidelberg, Königstuhl, D 69117 Heidelberg, Germany \label{I21}
\and Université Bordeaux, CNRS, LP2I Bordeaux, UMR 5797, F-33170 Gradignan, France \label{I22}
\and Institut für Astronomie und Astrophysik, Universität Tübingen, Sand 1, D 72076 Tübingen, Germany \label{I23}
\and Universität Innsbruck, Institut für Astro- und Teilchenphysik, Technikerstraße 25, 6020 Innsbruck, Austria \label{I24}
\and Universität Hamburg, Institut für Experimentalphysik, Luruper Chaussee 149, D 22761 Hamburg, Germany \label{I25}
\and Obserwatorium Astronomiczne, Uniwersytet Jagielloński, ul. Orla 171, 30-244 Kraków, Poland \label{I26}
\and Institute of Astronomy, Faculty of Physics, Astronomy and Informatics, Nicolaus Copernicus University, Grudziadzka 5, 87-100 Torun, Poland \label{I27}
\and Laboratoire Univers et Particules de Montpellier, Université Montpellier, CNRS/IN2P3, CC 72, Place Eugène Bataillon, F-34095 Montpellier Cedex 5, France \label{I28}
\and  INAF - Osservatorio Astronomico di Cagliari, Via della Scienza 5, I-09047 Selargius (CA), Italy \label{l29}
\and Nicolaus Copernicus Astronomical Center, Polish Academy of Sciences, ul. Bartycka 18, 00-716 Warsaw, Poland \label{I30}
\and University of Leicester, School of Physics and Astronomy, University Road, Leicester, LE1 7RH, United Kingdom \label{I31}
\and Fakult{\"a}t f{\"u}r Physik und Astronomie, Julius-Maximilians-Universit{\"a}t W{\"u}rzburg, Emil-Fischer-Str. 31, 97074, W{\"u}rzburg, Germany \label{WU}
\and Yerevan Physics Institute, 2 Alikhanian Brothers St., 0036 Yerevan, Armenia \label{I32}
\and Department of Physics, Konan University, 8-9-1 Okamoto, Higashinada, Kobe, Hyogo 658-8501, Japan \label{I33}
\and Kapteyn Astronomical Institute, University of Groningen, Landleven 12, 9747 AD Groningen, The Netherlands \label{I34}
\and GRAPPA, Anton Pannekoek Institute for Astronomy, University of Amsterdam, Science Park 904, 1098 XH Amsterdam, The Netherlands \label{I35}
}

\offprints{H.E.S.S.~collaboration,
\protect\\\email{contact.hess@hess-experiment.eu};
\protect\\\protect\footnotemark[1] Corresponding authors
\protect\\\protect\footnotemark[2] The authors honor the memory of Sergio Colafrancesco, who brought us together in pursuit of this study during his final years.
}

\abstract
{Composite galaxies that contain both Seyfert and starburst components may produce
very high-energy (VHE; $>100$~GeV) $\gamma$-ray emission at a wide range of spatial scales,
from a few Schwarzschild radii of a supermassive black hole (SMBH; $R_{\mathrm{S}}=10^{-6}$~pc for $M_{\mathrm{SMBH}}=10^7$~$M_{\odot}$) to dimensions of kiloparsec-size jet-driven outflows.
In addition to
supernova remnants, various sources have been suggested to explain data collected on composite galaxies, including multi-messenger
neutrino and ultra-high-energy cosmic-ray data.}
{The closest composite Seyfert–starburst galaxies (NGC 1068, the Circinus galaxy, and NGC 4945) are observed with
the High Energy Stereoscopic System (H.E.S.S.) to provide  constraints on cosmic-ray populations in these systems.}
{Data obtained in H.E.S.S. observations have been analyzed to search for VHE $\gamma$-ray counterparts to the GeV $\gamma$-ray signals detected with \textit{Fermi}-LAT and for potential spectral components in the VHE range.}
{No significant signals have been found in these H.E.S.S. data. Upper limits on the VHE $\gamma$-ray fluxes were applied to constrain theoretical models involving different spectral components.
}
{}

\keywords{Gamma rays: galaxies -- Galaxies: Seyfert -- Galaxies: starburst -- ISM: jets and outflows}

\maketitle

%LEFT COLUMN
%
%\newpage
%
%RIGHT COLUMN
%
%\newpage

\section{Introduction}

Seyfert galaxies are mostly spiral galaxies that host a supermassive black hole (SMBH) powering an active galactic nucleus \citep[AGN; for a review, see][]{Begelman1984}. 
In contrast, the Milky Way---a spiral galaxy hosting an SMBH \citep[][]{Ghez2008, Gillessen2009}---currently does
not have an AGN.
Large-scale bubbles observed in the radio band in Seyfert galaxies,
including the Circinus galaxy \citep[][]{ReferenceElmouttie, ReferenceEbrahim} and NGC 1068 \citep[][]{ReferenceWilson}, bear a strong resemblance to those blown out of the Milky Way.
The latter include the \textit{Fermi} and \textit{eROSITA} bubbles, which are kiloparsec-scale $\gamma$-ray and X-ray emitting lobes \citep[][]{ReferenceSu, Predehl2020}. They are thought to have been created by a past episode of Seyfert-like activity lasting for $10^{5-6}$ years at the Galactic center \citep[e.g.,][]{Guo2012, Zubovas2012}.
In addition, 430-parsec bipolar radio bubbles in the Galactic center revealed by MeerKAT \citep[][]{Heywood2019} may be a less energetic
version of the process that created the \textit{Fermi} bubbles.
Reflection of X-rays from Sgr A$^{\ast}$ by dense gas in the Galactic center region allows the study of past flaring activity of the
SMBH over hundreds to thousands of years \citep[][]{Vainshtein1980}.
Polarized X-ray emission from molecular clouds in the Galactic center implies that 200 years ago, the X-ray luminosity of Sgr A$^{\ast}$ was briefly comparable to that of a Seyfert galaxy \citep[][]{Marin2023}.
Research on Seyfert galaxies may provide insight into the analogy between bubble-like structures in Seyfert galaxies and the Milky Way.

Models of AGN invoke an SMBH to supply the power that efficiently turns into radiation \citep[e.g.,][]{Blandford1976}. The classes of
AGN are organized according to their radio-loudness.
Radio-loud AGN, including blazars and radio galaxies, are well-known sources of
very high-energy (VHE; $>100$~GeV) $\gamma$-ray emission \cite[][]{Punch1992, RefM87, CenA2009}.
The emission from blazar jets is greatly enhanced by the effect of Doppler boosting due
to their jets oriented close to the line of sight. 
Seyfert galaxies host radio-quiet AGN, which have much less powerful jets or no jets at all. The dimming in emission from the outflows in Seyfert galaxies (due to an intrinsically low luminosity) and in radio galaxies (due to an orientation at large angles to our line of sight) brings a new perspective on examining otherwise hidden emission components.
If pions are created in plasma particle interactions that involve sufficiently energetic hadrons, then neutral pion decays give rise to VHE $\gamma$ rays, while charged
pion decays result in VHE neutrinos. This makes the VHE $\gamma$-ray/neutrino relationship a viable way of probing the innermost regions of AGN
\citep[e.g.,][]{Eichler1979}.
VHE $\gamma$-ray observations of nearby Seyfert galaxies, such as NGC 1068,
provide information necessary for {a joint study} of the propagation of VHE neutrinos and hadronic
$\gamma$ rays for a black hole 
surrounded by gas or photons. Models differ in the predictions of the ratio of $\gamma$-ray to neutrino fluxes due to the opacity of $\gamma$ rays and,
as a consequence, neutrino emission can be accompanied by a very small $\gamma$-ray flux \citep[][]{ReferenceBerezinsky}.
The Seyfert galaxy NGC 1068 is one of a few prominent VHE neutrino source candidates identified by the IceCube collaboration \citep[][]
{ReferenceIcecubeNGC1068excess}.
However, no VHE $\gamma$ rays were detected from NGC 1068 in  H.E.S.S.
observations with low exposure \citep[][]{HESSupperLimits1, HESSupperLimits} and in MAGIC observations with deep
exposure of 125 hours \citep[][]{ReferenceMAGIC}.
The model by \citet[][]{ReferenceKheirandish} suggests that the Circinus galaxy is the next most promising VHE neutrino emitter among Seyfert galaxies.

Starburst galaxies are characterized by higher star formation rates than regular spiral galaxies.
The two archetypical starburst galaxies NGC 253 and M82 were discovered to emit VHE $\gamma$ rays \citep[][]{ReferenceHESS1,
ReferenceVERITAS}.
NGC 1068, the Circinus galaxy, and NGC 4945 are composite Seyfert--starburst galaxies having strong starburst activity
in addition to a Seyfert nucleus. Their $\gamma$-ray emission, which was detected with \textit{Fermi}-LAT, may come from galactic disks and/or AGN \citep[][]{ReferenceLenain, ReferenceHayashida}.
NGC 4945 is believed to be
one of the likely ultra-high-energy cosmic-ray (UHECR) sources along with another starburst galaxy M82 based on
the results by the Pierre Auger Observatory (PAO) and the Telescope Array experiment \citep[][]{ReferenceAuger, ReferenceAuger2022,diMatteo2023}.

This paper presents the results of deep observations of NGC 1068 and observations of the Circinus galaxy and NGC 4945 with H.E.S.S.
The paper is organized as follows. The properties of composite galaxies and their potential for  detection in VHE $\gamma$ rays are provided in Sect. \ref{sect:galaxies}. The data analysis and results are reported in Sect. \ref{sect:obs}.
The results are compared with predictions in Sect. \ref{sec4}. A summary is presented in Sect. \ref{sect:sum}.

\section{Composite galaxies as potential VHE sources}
\label{sect:galaxies}

NGC 253 is a faint VHE source \cite[][]
{RefNGC253revisited},  and its detection with H.E.S.S. at a 5~$\sigma$ level required a very long exposure of 119 hours \cite[][]
{ReferenceHESS1}. NGC 253, the Circinus galaxy, and NGC 4945, are located at comparable distances. Therefore, in the absence of relatively strong $\gamma$-ray emission from AGN outflows in the latter two galaxies, their detection may similarly require very long exposures. The high-energy (HE; from $100$~MeV to $100$~GeV) $\gamma$-ray luminosity of NGC 1068
in the GeV band is an order of magnitude higher than that of the Circinus galaxy and NGC 4945, while its distance is about four times greater. Consequently, detecting VHE $\gamma$-ray emission
from NGC 1068 may also necessitate deep observations \cite[see][]{ReferenceMAGIC}.
Notably, models for $\gamma$-ray emission from Seyfert galaxies offer significant flexibility, allowing for VHE $\gamma$-ray fluxes detectable by H.E.S.S.

\subsection{Composite galaxies: NGC 1068, Circinus, and NGC 4945}

NGC 1068 is a face-on, early type, barred spiral galaxy located near the celestial equator and is the brightest Seyfert 2
galaxy \citep[][]{deVaucouleurs1973}. The distance to NGC 1068 remains somewhat uncertain. The literature describes several
measurements, including $10.1\pm1.8$~Mpc \citep[][]{Tully2009},
$11.1\pm0.5$~Mpc \citep[][]
{Tikhonov2021}, $14.0\pm2.1$~Mpc \citep[][]{Anand2021}, and $16.7\pm3.0$~Mpc \citep[][1$^{\prime}$ corresponds to $4.9$~kpc at this distance]{Gao2004}.
For consistency with \citet[][]{FermiSFG2012} and
\citet{ReferenceWang}, the distance of $16.7$~Mpc is adopted in this paper.
NGC 1068 exhibits vigorous starburst activity. It hosts a circumnuclear disk about $200$~pc in radius,
surrounded by a $2$~kpc starburst ring connected to the circumnuclear disk by a bar \citep[][]{Garcia2014}.
It also harbors a bright radio source with a prominent radio jet \citep[][]{Muxlow1996} and extended radio lobes \citep[][]{ReferenceWilson}.
The mass of the SMBH in NGC 1068 is uncertain, with estimates ranging from $8\times10^{6}$~$M_{\odot}$ to $1.5\times10^{7}$~$M_{\odot}$ \citep[e.g.,][]{Greenhill1997, Hure2002, Lodato2003}.

The Circinus galaxy, at a distance of only $4.2\pm0.7$~Mpc
\citep[][1$^{\prime}$ corresponds to $1.2$~kpc at this distance]{Tully2009}, is a gas-rich spiral galaxy in
the southern constellation of Circinus. It is located about 4$^{\circ}$ below the Galactic plane and was not discovered until
1977 \citep[][]{Freeman1977}. Much of the gas and dust in the disk of the Circinus galaxy is concentrated in two rings, which are 80
and $430$~parsec accross, surrounding its center, and hosting an immense starburst region \citep[e.g.,][]
{Wilson2000}. At the center there is a Seyfert 2 nucleus, the signature of an SMBH that is accreting surrounding gas and dust.
The galaxy shows a spectacular, one-sided [O III] ionization cone
\citep[][]{Marconi1994}, whose asymmetry is probably due to extinction by the Circinus galaxy disk (an inclination angle of
$i\simeq65^{\circ}$).
Additionally, it exhibits bipolar radio lobes, each with a size of $\sim1.5$~kpc, inflated by kiloparsec-scale outflows \citep[][]{ReferenceElmouttie, ReferenceEbrahim}.

NGC 4945, at a distance of only $3.8\pm0.3$~Mpc \citep[][1$^{\prime}$ corresponds to $1.1$~kpc at this distance]{Karachentsev2007},
is a barred spiral galaxy seen
almost edge-on ($i\simeq80^{\circ}$) and located in the southern constellation of Centaurus. It is among the closest objects
where AGN and starburst activity coexist. Its SMBH mass is around $\sim10^{6}$~$M_{\odot}$, close to that of our own Galaxy, but
accreting at a much higher rate. 
Similar to the Circinus galaxy, NGC 4945 contains a highly obscured Seyfert 2 nucleus \citep[][]{Iwasawa1993}.
Its nucleus is a hard X-ray source variable on timescales of hours \citep[][]{Puccetti2014} and one of the
brightest Seyfert 2 AGN at energies above 20 keV \citep[][]{Itoh2008}. Below 10 keV, its X-ray emission is only visible through reflected
emission due to the large column density ($\mathrm{log}[{N}_\mathrm{H}/{\mathrm{cm}}^{-2}]\sim 24.7$) that completely
blocks the primary nuclear emission. The nucleus is surrounded by a starburst disk with $200$~pc (11$^{\prime\prime}$) diameter
oriented in the same direction as the galactic disk \citep[][]{Marconi2000}.

\subsection{VHE flux expectations based on GeV emission}

NGC 1068 and NGC 4945 were identified with \textit{Fermi}-LAT sources based on 1.6 years of data, while the
identification of the Circinus galaxy required 4 years of data \citep[][]{ReferenceLenain, ReferenceHayashida}.
Since then, \textit{Fermi}-LAT has continuously been collecting data and significantly increased the number of $\gamma$ rays detected from
these sources. The flux normalizations and the photon indices for NGC 1068 and NGC 4945 reported by \citet[][]{ReferenceLenain}
are compatible with the values reported in the \textit{Fermi}-LAT 14-year catalog of $\gamma$-ray sources \citep[4FGL-DR4;\footnote{\tiny{\url{https://fermi.gsfc.nasa.gov/ssc/data/access/lat/14yr_catalog/}}}][]{4FGLDR3} within statistical errors.
As for the Circinus galaxy, the flux normalization (see Table \ref{tab:circ}) reported by \citet[][]{ReferenceGuo} and
\citet[][]{ReferenceEbrahim}, and listed in the 4FGL-DR4 catalog is lower than that initially reported by \citet[][]{ReferenceHayashida}.
This difference in the GeV flux of the Circinus
galaxy results in a longer exposure time needed to detect this source with H.E.S.S. The HE $\gamma$-ray photon indices of NGC 1068, the Circinus galaxy, and NGC 4945 are $2.31\pm0.13$, $2.20\pm0.14$, and $2.31\pm0.10$ \citep[][]{ReferenceLenain, ReferenceGuo} and systematically harder than that of the Galactic diffuse emission \citep[][]{diffuse2012}.
Hereafter, this paper adopts the 4FGL-DR4 catalog's Fermi-LAT results \citep[][]{4FGLDR3} as a benchmark for estimating expectations and for reporting in the figures and compares it to published results from other works where appropriate.

\begin{table}
\centering
\caption{Gamma-ray flux from the Circinus galaxy measured with \textit{Fermi}-LAT.}
\begin{tabular}{|c | c |}
 \hline
 Flux ($>0.1$~GeV) & Reference \\
 $10^{-9}$~ph cm$^{-2}$s$^{-1}$ & \\
 \hline
$18.8\pm5.8$ & \citet[][]{ReferenceHayashida} \\
$5.7\pm2.0$ &  \citet[][]{ReferenceGuo} \\
$6.1\pm2.0$ & \citet[][]{ReferenceEbrahim}  \\
 \hline
\end{tabular}
\label{tab:circ}
\end{table}

Exposure times necessary for 5 $\sigma$ detections in the VHE $\gamma$-ray band are estimated on the basis of
the sensitivity of the H.E.S.S. array observing at a zenith angle of 18$^{\circ}$ (for NGC 1068
and NGC 4945) or 37$^{\circ}$ (for the Circinus galaxy).
If adopting the best-fit spectral parameters from the 4FGL-DR4 catalog, the corresponding exposure times for NGC 1068, the Circinus galaxy, and NGC 4945 are 342 hrs, 98 hrs, and 64 hrs.
The lack of evidence in the 4FGL-DR4 catalog that these HE $\gamma$-ray sources are variable makes the extrapolation based on the square-root-of-exposure-time rule valid (for daily variability of NGC 4945, see Sect. \ref{sect:temp}).
Given that the single broad spectral component commonly softens from the GeV to the VHE band, the extrapolation based on a phenomenological power-law model is optimistic for such a component; even this approach yields long exposure times for potential detections of these three targets.

Extrapolation of the spectrum from GeV to TeV energies leads to an uncertainty in differential TeV flux because of
the statistical error on the photon index in the GeV band.
The VHE $\gamma$-ray excesses predicted for the H.E.S.S. observations take this uncertainty into account and are reported in Sect. \ref{sect:results}. Moreover, the photon index in the GeV band can differ from that in the VHE band
in the presence of two or more spectral components. H.E.S.S. observations reported in Sect. \ref{sect:obs} allowed us to assess VHE properties of these three composite galaxies and to search for new spectral components.

\section{H.E.S.S. observations and results}
\label{sect:obs}

This section reports the results obtained from observations of three composite Seyfert/starburst galaxies with H.E.S.S.

\subsection{H.E.S.S. experiment and analysis methods}

The H.E.S.S. array consists of five imaging atmospheric Cherenkov telescopes (IACTs). It is located in Namibia
at 23$^{\circ}$16$^{\prime}
$18$^{\prime\prime}$ southern latitude, 16$^{\circ}$30$^{\prime}$00$^{\prime\prime}$ eastern longitude at an altitude of 1800
m above sea level \citep[][]{HESS2006}.
Originally
consisting of four 12-meter telescopes (CT1-4), the array was expanded in 2012, becoming HESS-II with the addition of a fifth larger
28-meter telescope (CT5). The 12-meter telescopes, arranged in a square with 120-m sides, have been in operation since 2004 \citep[][]{Hinton2004}. H.E.S.S. employs the stereoscopic imaging atmospheric Cherenkov technique \citep[e.g.,][]{Daum1997}.
The array with the four 12-meter telescopes is sensitive to $\gamma$ rays above an energy threshold $\sim$0.1 TeV for observations at zenith,
up to energies of tens of TeV. The 28-meter telescope is located at the center of the array and potentially extends the energy
range covered by the array down to energies of $\sim20$~GeV. The H.E.S.S. II experiment is the first hybrid Cherenkov instrument.
The observations with H.E.S.S. reported in this paper were performed in so-called wobble mode \citep[][]{Fomin1994, Berge2007}. In this mode, the source is alternatively offset from the pointing
direction in Right Ascension and Declination typically by $0\fdg7$. The data were recorded in 28-minute exposures, called observation runs, to minimize systematic
changes in instrumental response. To reduce systematic effects arising from imperfect instrument or atmospheric conditions, good-quality runs \citep[][]{HESS2006} were selected.

The analyses presented in this paper are based on a semi-analytical model of air showers for the event reconstruction and 
background suppression \citep[][]{ModelAnalysis2009}. This \texttt{model analysis} provides an improved angular resolution and a better sensitivity compared with the traditional image-moments-fitting (Hillas parameter-based) analysis. The gain in
sensitivity makes the \texttt{model analysis} powerful for studying faint VHE $\gamma$-ray sources.
The cut configuration requiring a minimum of 60 photoelectrons per image was used.
The On-source counts were extracted from the circular region of $0\fdg12$ radius centered on each of the three galaxies. This
radius corresponds to the selection cut for a VHE point-like source. Each of these galaxies is considered as a point-like
source for H.E.S.S. The reflected region background method with multiple Off-source regions \citep[][]{Berge2007} was used to measure the background for $\gamma$-ray-like events. The source significance was calculated using Eq. (17) from \citet[][]
{LiMa1983}. Following the method of \citet[][]{Rolke2005}, differential flux upper limits (ULs) were derived at the 95\% confidence
level under the assumption of a power-law spectrum,\footnote{These ULs are not very sensitive to the choice of the photon index.} $dN/dE\propto E^{-\Gamma}$, with index $\Gamma$=2.4.
The results reported in this paper were cross-checked with an independent analysis performed with gammapy \citep[][]{gammapy2023}, by applying the
high-level analysis pipeline to data that are processed with an independent low-level chain \citep[ImPACT;][]{Parsons2014}.

\subsection{Observations, data analyses, and results}
\label{sect:results}

This section is divided into three parts focusing on NGC 1068, the Circinus galaxy, and NGC 4945. In each subsection, a summary of the performed H.E.S.S. observations is provided, along with their results.

\begin{table}
\centering
\caption{H.E.S.S. analysis results.}
\begin{tabular}{|c | c | c | c | c | c |}
 \hline
 Data set &  On & Off & $\alpha$ & Excess & Signifi- \\
      & (cts) & (cts) & & (cts) & cance ($\sigma$) \\
 \hline
 NGC 1068 & 6077 & 77660 & 0.0762 & 156.7 & 2.0 \\
 Circinus & 213 & 2923 & 0.06075 & 35.4 & 2.5 \\
 NGC 4945 & 622 & 7460 & 0.0821 & 9.4 & 0.4 \\
 \hline
\end{tabular}
\tablefoot{The first column represents the data set. The second and third
columns show the number of signal plus background events around the source position and background events from the off-source region, respectively. The fourth column shows background normalization $\alpha$. The fifth and sixth columns show the $\gamma$-ray excess and significance, respectively.}
\label{ana}
\end{table}

\subsubsection{NGC 1068}

H.E.S.S. observations of NGC 1068 were performed from 2004 to 2016.
A total of 431 good-quality runs were recorded in the wobble mode.
Among these, 354 runs were taken with 12-meter telescopes, 23 runs with only the 28-meter telescope, and 54 runs employed 12-meter telescopes and the 28-meter telescope.
Given that the MAGIC observations of NGC 1068 started in
January 2016, 279 of the 431 runs, that is 65\%, were accumulated with H.E.S.S. prior to the start of MAGIC observations. This 
percentage increases to 75\% if only observations conducted with 12-meter telescopes are analyzed.
The zenith angle for the observations is in the 
range from 23$^{\circ}$ to 38$^{\circ}$, and the mean zenith angle is 26$^{\circ}$.
They were mostly taken with an offset of $0\fdg7$ from the target direction, that is (RA, Dec)=(40\fdg670, -0\fdg013).
The total exposure is 168.5 hours corresponding to an exposure of 144.3 hours corrected for the camera acceptance depending on off-optical axis angle.

The analysis of the H.E.S.S. data on NGC 1068 yields a $\gamma$-ray excess of 157 counts above the background (Table \ref{ana}). The excess corresponds to a significance of $2.0 \sigma$
and hence no significant signal has been found.
Since no significant signal was detected, upper limits were derived. The differential flux upper limits at a confidence level of 95\% are shown in Fig. \ref{plot_ngc1068} for the full H.E.S.S. data set.
This figure along with the other figures in this paper also shows the \textit{Fermi}-LAT bow tie and the \textit{Fermi}-LAT 0.1-1 TeV spectral point based on 14 years of \textit{Fermi}-LAT data and taken from the 4FGL-DR4 catalog \citep[][]{4FGLDR3}.
The LAT spectral point taken from 4FGL-DR4 is consistent with that presented by
\citet[][]{Ajello2023},
but reported at a marginally higher significance level of $\sqrt{TS}=3.4$ vs $\sqrt{TS}=2.8$, where $TS$ is a test statistic \citep[][]{Mattox}, and with smaller uncertainties.
\footnote{The LAT spectral points taken from 4FGL-DR4 and shown in Figs. \ref{plot_circinus} and \ref{plot_ngc4945} are consistent with the results from \citet[][]{Murase2024},
but obtained at a marginally higher significance level of $\sqrt{TS}=2.2$ vs $\sqrt{TS}=1.9$ for the Circinus galaxy and $\sqrt{TS}=2.9$ vs $\sqrt{TS}=2.8$ for NGC 4945.}
In addition, to increase the fraction of data taken prior to the start of MAGIC observations of NGC 1068,
an analysis of observations conducted with only the 12-meter telescopes was performed.
This results in an excess of 55.8 events corresponding to a significance of $1.8 \sigma$. This suggests that NGC 1068 did not experience any strong, temporal VHE flux variability.
Overall, the flux upper limits derived from the deep H.E.S.S. observations are as tight as those derived from the deep
MAGIC observations.

Based on 4FGL-DR4, the VHE $\gamma$-ray excess expected for the H.E.S.S. observations of NGC 1068 is at a statistical significance ranging from 2.1$\sigma$ to 4.7$\sigma$. The non-detection of NGC 1068 with H.E.S.S. is, thus, consistent with the expectation based on the \textit{Fermi}-LAT flux extrapolation within the framework of a single-power-law model.

\subsubsection{Circinus galaxy}

H.E.S.S. observations dedicated to the Circinus galaxy were performed from 1st of March 2014 to 5th of May 2014. A total of
26 good-quality runs were recorded in the wobble mode.
Among these, 21 runs were taken with at least four telescopes, including 28-meter telescope, and 5 runs with 12-meter telescopes.
The zenith angle for the observations in 2014 is in the range from 42\fdg1 to 44\fdg6.
The offset from the target direction, that is (RA, Dec)=(213\fdg292, -65\fdg343), for these observations is $0\fdg5$. Alongside the 26 dedicated observation runs on Circinus, an additional 18 runs targeting nearby sources were also included.
The latter consist of 12 runs taken in May 2006 in the direction of the
pulsar PSR J1357-6429, and 6 runs taken in March 2007 in the direction of the pulsar wind nebula HESS J1356-645. Despite the
$1\fdg9$ separation between the Circinus galaxy and PSR J1357-6429, some of these observations in wobble mode
provide a smaller distance, as close as $1\fdg3$, between the camera center and the Circinus galaxy.
The additional runs correspond to H.E.S.S. data collected with 12-meter telescopes.
The cumulative exposure resulting from all
these observations is about 18.4 hours corresponding to an acceptance-corrected exposure of 13.2 hours.

Among the three Seyfert galaxies discussed here, the Circinus galaxy is the closest to the Galactic plane, $\mathrm{b}
=-3\fdg80$. At this Galactic latitude, the Galactic diffuse $\gamma$-ray background is still a subdominant
component of the background of $\gamma$-ray-like events for H.E.S.S. To prevent background-signal contamination from a known VHE $
\gamma$-ray source in the field of view, HESS J1356-645, an exclusion region was defined around this source \citep[][]
{HESSJ1356}.
The analysis of the H.E.S.S. data accumulated toward the Circinus galaxy yields a $\gamma$-ray excess of 213 $\gamma$-ray-like events above the background corresponding to a significance of $2.5 \sigma$ (Table \ref{ana}).
No significant signal has been found. The differential flux upper limits at a confidence level of 95\% are shown in Fig. \ref{plot_circinus}.

Based on 4FGL-DR4, the expected VHE $\gamma$-ray excess for the H.E.S.S. observations of the Circinus galaxy ranges from 0.8$\sigma$ to 2.8$\sigma$ in significance. Therefore, the non-detection of the Circinus galaxy by H.E.S.S. is consistent with the expectation derived from the \textit{Fermi}-LAT flux extrapolation under the assumption of a single-power-law model.

\subsubsection{NGC 4945}

H.E.S.S. observations of NGC 4945 were performed between 2012 and 2015.
A total of 104 good-quality runs were recorded in the wobble mode.
Among these, 44 runs were taken with 12-meter telescopes and 60 runs in 2015 with only the 28-meter telescope.
The zenith angle for the observations with 12-meter telescopes is in the range from
26\fdg3 to 33\fdg4.
The offset from the pointing direction, that is (RA, Dec)=(196\fdg365, -49\fdg468), for the observations with 12-meter telescopes is 
$0\fdg5$ and the offset for the observations with the 28-meter telescope is $0\fdg7$. The total exposure resulting from all these
observations is 42.7 hours corresponding to an acceptance-corrected exposure of 37.2 hours. The acceptance-corrected exposure
accumulated with only the 12-meter telescopes is 17.5 hours and that with the 28-meter telescope is 20.7 hours. 92\% and 63\% of the runs
taken with the 12-meter telescopes and with the 28-meter telescope, respectively, were accumulated when the Swift-BAT count rate 
of NGC 4945 in the 15–50 keV range
was above $1.71\times10^{-3}$~cts~cm$^{-2}$~s$^{-1}$. This count rate defines a high hard X-ray state \citep[][]{ReferenceWojaczynski}.

The analysis of the H.E.S.S. data accumulated toward NGC 4945 yields a $\gamma$-ray excess of 9.4 $\gamma$-ray-like events above the background corresponding to a significance of $0.4 \sigma$ (Table \ref{ana}).
Given the nearly equal division of the total exposure between observations with the 12-meter telescopes and with the 28-meter telescope, each of these two data subsets was also analyzed separately.
The analysis of data from the 12-meter telescopes yields a $\gamma$-ray excess of 17.8 events corresponding to a significance of $0.9 \sigma$, while the analysis of data from the 28-meter telescope yields no excess.
No strong excess is observed in any of these sets of data and flux upper limits were derived at a 95\% confidence level.
The differential flux upper limits obtained from H.E.S.S. observations of NGC 4945 with the 12-meter telescopes are 
shown in Fig. \ref{plot_ngc4945}. 

Based on 4FGL-DR4, the VHE $\gamma$-ray excess expected for the H.E.S.S. observations of NGC 4945 with 12-meter telescopes is at a significance ranging from 1.9$\sigma$ to 3.3$\sigma$. The non-detection of NGC 4945 with H.E.S.S. is consistent with the expectation based on the \textit{Fermi}-LAT flux extrapolation using a single-power-law model. Since the obtained significance value ($0.9\sigma$) is lower than 3.3$\sigma$, it suggests that the photon index is softer than the lower bound of 2.2 of the $68\%$ confidence interval from 4FGL-DR4.

To check whether VHE $\gamma$-ray emission is correlated with hard X-ray emission, the analyses of H.E.S.S. data recorded during high and low hard X-ray states were performed separately. None of these analyses reveals a significant excess.

\section{Discussion}
\label{sec4}

Over the last decade and a half, $\gamma$-ray astronomy has been enriched by observations in the HE $\gamma$-ray band with \textit{Fermi}-LAT and the largest neutrino and UHECR
observatories to date, IceCube and PAO,
have been operating. Gamma-ray emission from starburst galaxies, composite galaxies, and the \textit{Fermi} bubbles in the Milky Way was discovered by \textit{Fermi}-LAT, enhancing our knowledge of galactic emission components.
The detection of astrophysical neutrinos made by the IceCube neutrino observatory \citep[][]{IceCube1} opened a new window to study HE astrophysical sources via the multi-messenger approach. NGC 1068, the blazar TXS 0506+056, and the Milky Way Galactic plane have been shown to be likely VHE neutrino
sources \citep[][]{IceCube2, IceCube3, IceCubeGalPlane}. Another multi-messenger channel relevant for these composite galaxies is
UHECRs.
In contrast to the isotropic model of UHECRs, their origin from starburst galaxies
is favored by the PAO data with $4.0\sigma$ confidence \citep[][]{ReferenceAuger}. However, other models, including the
AGN model of UHECR origin, remain possible.

Together with $\gamma$ rays, neutrino and UHECR messengers have led to renewed interest in theoretical models
describing the observed properties of composite Seyfert/starburst galaxies.
Among these properties are the $\gamma$-ray flux level, as well as the $\gamma$-ray spectral and temporal characteristics.
The H.E.S.S. observations of NGC 1068, the Circinus galaxy, and NGC 4945 described in the previous section resulted in flux upper limits on VHE $\gamma$-ray emission.
To put these upper limits into a larger context, below they are compared with the fluxes obtained with other instruments and
with the expectations from theoretical models.

\subsection{VHE $\gamma$-ray emission due to AGN outflows}

Gamma-ray emission produced by AGN outflows can have a spectral shape different from a single power law and/or be variable in 
time. Models involving AGN have been proposed to describe the measured HE $\gamma$-ray fluxes. Such models can produce signals in the
TeV $\gamma$-ray band of lower or higher luminosity than those expected from a single power-law extrapolation from the GeV $\gamma$-ray
band to the TeV $\gamma$-ray band. On the one hand, in the external inverse Compton (EIC) model \citep[][]{ReferenceLenain}, VHE $\gamma$-ray
emission can be within reach of H.E.S.S. only if the maximum energy
of electrons is sufficiently high, $\gamma_\mathrm{max}>5\times10^6$. However, given that the characteristic frequency of a
soft (external) photon field for NGC 1068 is about $10^{14}$ Hz, the electrons that produce VHE emission up-scatter the soft
photon targets in the Klein-Nishina regime. Because of the Klein-Nishina suppression of EIC emission at the highest energies,
the VHE $\gamma$-ray emission predicted by the EIC model remains small. On the other hand, an additional $\gamma$-ray component
with a hard spectrum, $\Gamma\simeq2.0$, such as that introduced by \citet[][]
{ReferenceLamastra} can account for $\gamma$-ray emission due to protons accelerated by AGN-driven shocks.
Their AGN outflow model, however, requires 
acceleration efficiencies for cosmic-ray (CR) protons higher than those commonly assumed in supernova remnant (SNR) shocks. The constraints on parameters of
the AGN outflow model derived by \citet[][]{ReferenceMAGIC} showed that this model fails to reproduce the broadband $\gamma$-ray emission from NGC 1068. Sections \ref{bubbles} and \ref{sect:temp} describe models of AGN jet activity constrained by H.E.S.S. observations.

\subsubsection{Jet-driven bubbles: NGC 1068 and the Circinus galaxy}

\label{bubbles}

The \textit{Fermi}-LAT discovery of the \textit{Fermi} bubbles
\citep[][]{ReferenceSu} suggests that similar $\gamma$-ray-emitting
structures can exist in other galaxies. The kiloparsec-scale lobes emitting in the radio band in NGC 1068 \citep[][]
{ReferenceWilson} and the Circinus galaxy \citep[][]{ReferenceElmouttie, ReferenceEbrahim} are candidates.
To account for the different astrophysical environments in the Milky Way, NGC 1068, and the Circinus galaxy, the model for the
HE $\gamma$-ray emission from the \textit{Fermi} bubbles by \citet[][]{ReferenceKataoka} requires adjustments. The original model by \citet[][]{ReferenceKataoka} is a one-zone leptonic model in which the radio emission and the $\gamma$-ray emission
arise from the same population of relativistic electrons. These emissions occur through the synchrotron process and the inverse-Compton scattering of
cosmic-microwave-background photons, respectively. The model assumes that the electron spectrum is a broken power-law function
and that the difference between spectral indices above and below the break equals 1.
The first adjustment to the original model is that in the more generalized version of the model the
interstellar radiation field is considered in addition to the cosmic microwave background field. This difference is
significant because the bubbles in NGC 1068 and the Circinus galaxy are located closer to the galactic disks, and the infrared luminosity
of NGC 1068 is much stronger than that of the Milky Way. Moreover, on average, photons of the interstellar radiation field are up-scattered
by relativistic electrons
to higher energies than the cosmic microwave background field. The model of the interstellar radiation field is adopted from the GALPROP code \citep[][]
{ReferencePorter, ReferenceStrong} and scaled according to the galactic infrared luminosity.
This scaling is supported because the radiation field energy density in starburst galaxies exceeds that of the Milky Way \citep[e.g.,][]{ReferenceOhm, ReferenceYoast2}.
In our leptonic model, $
\gamma$-ray emission is luminous at TeV energies but relatively faint in the GeV band, similar to the leptonic models of young
shell-type TeV-bright SNRs \citep[e.g.,][]{Condon2017}.

\begin{figure}
\includegraphics[width=0.5\textwidth]{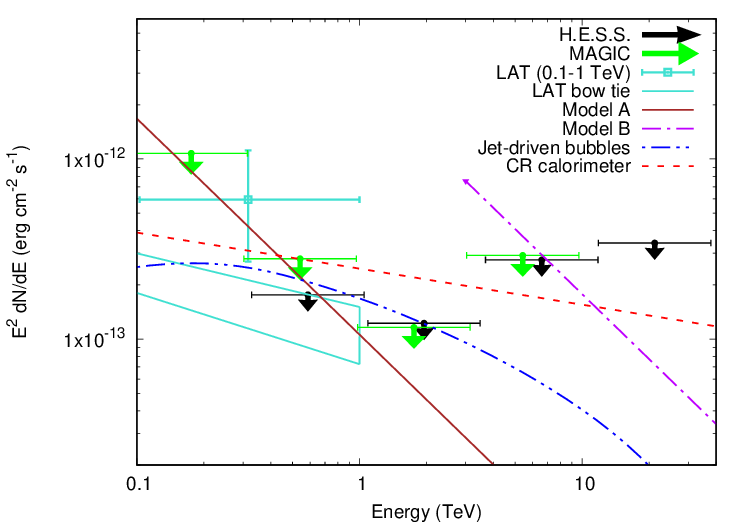}
\caption{Flux upper limits obtained from H.E.S.S. observations of NGC 1068.
This figure also shows MAGIC flux upper limits \citep[][]{ReferenceMAGIC}
along with a \textit{Fermi}-LAT data bow tie and an 0.1-1 TeV data point from the
\textit{Fermi}-LAT catalog \citep[][]{4FGLDR3}.
Models A and B show the absorbed $\gamma$-ray flux from the neutrino source (see Sect. \ref{sect_neutrino}). A model for jet-driven bubbles (Sect. \ref{bubbles}) is provided. The calorimetric bound is also shown.}
\label{plot_ngc1068}
\end{figure}

\begin{figure}
\includegraphics[width=0.5\textwidth]{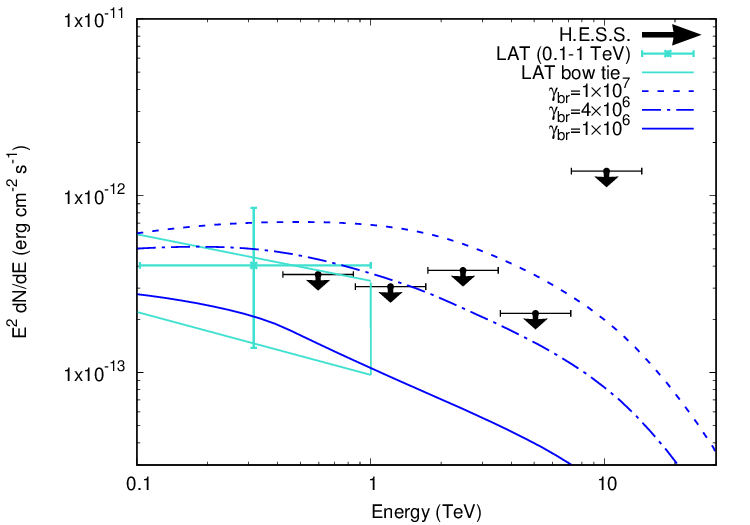}
\caption{Flux upper limits from H.E.S.S. observations of the Circinus galaxy. 
This figure also shows a \textit{Fermi}-LAT data bow tie and an 0.1-1 TeV data point from the \textit{Fermi}-LAT catalog \citep[][]{4FGLDR3}.
A model involving gamma-ray emitting lobes is shown for different values of a spectral break.}% (for more details, see Section \ref{bubbles})
\label{plot_circinus}
\end{figure}

The second adjustment to the original model is that CR electrons and magnetic fields are assumed to be in approximate energy
equilibrium. Under this assumption, the magnetic field strengths are estimated on the basis of the flux densities measured in the radio band. The magnetic field strength in each of the lobes of the Circinus
galaxy is about 3.5 $\mu$G, while the magnetic field strengths in the northeast and southwest lobes of NGC 1068 are 87 $
\mu$G and 44 $\mu$G, respectively.
For the lobes of the Circinus galaxy and NGC 1068, the total energy required in magnetic fields and relativistic electrons is $(2-4)\times10^{53}$ erg, which is a factor of more than $100$ smaller than that used by \citet[][]{ReferenceKataoka} for the Fermi bubbles. Given the typical kinetic power of jets in Seyfert galaxies of $10^{42}$ erg/s, such energy can be supplied in an activity episode lasting for $\sim10^4$ years.
The electron spectra are modeled as a broken power law with a spectral index $s=2.2$ below the
break energy and $s=3.2$ above the energy break.
The minimum and maximum Lorentz factors of the electrons are adopted from \citet[][]{ReferenceKataoka}. With these settings, the electron break
Lorentz factor is the only free parameter of this model.
The dash-double-dotted line in Fig. \ref{plot_ngc1068} illustrates the model of VHE emission from the lobes of NGC 1068, which is computed for an electron break Lorentz factor of $3\times10^{6}$.
Figure \ref{plot_circinus} shows the model of VHE emission from the lobes of the Circinus galaxy computed for different electron
break Lorentz factors and compared to the flux upper limits obtained from the H.E.S.S. observations.
In order not to overproduce the derived H.E.S.S. flux upper limits, the electron break Lorentz factor, $\gamma_{\mathrm{brk}}$, for NGC 1068 and
the Circinus galaxy must be lower than $3\times10^{6}$ and $4\times10^{6}$, respectively.
The best-fit value of the energy break obtained by \citet[][]{ReferenceKataoka} for the \textit{Fermi} bubbles is $
\gamma_{\mathrm{brk}}=10^{6}$ and below the upper limits on $\gamma_{\mathrm{brk}}$ for NGC 1068 and the Circinus galaxy.

The radiative loss rate is approximately 2.5 times higher in the Circinus lobes, and approximately 50 times higher in the northeastern (NE) lobe and 13 times higher in the southwestern (SW) lobe of NGC 1068 than in the Fermi bubbles for the same electron energy. Consequently, the electron cooling time is shorter by these same factors. The source age for the lobes of both the Circinus galaxy and NGC 1068 is shorter by a factor of 4 for the Circinus lobes, and by factors of 30 for the NE lobe and 18 for the SW lobe of NGC 1068 than that of the Fermi bubbles. This age difference results from their smaller radii -- 1 kpc for the Circinus lobes, and 0.13 kpc for the NE lobe and 0.22 kpc for the SW lobe of NGC 1068 -- compared to the approximately 4 kpc radius of the Fermi bubbles. Therefore, the energy break in the electron spectrum is expected to be only $\sim1.5$ times higher for the Circinus lobes and the SW lobe of NGC 1068, but $\sim1.5$ times lower for the NE lobe of NGC 1068, than in the Fermi bubbles.

\subsubsection{Temporal and spectral characteristics of NGC 4945}
\label{sect:temp}

In contrast to the jet-driven bubble model resulting in steady emission, there are models predicting variable $\gamma$-ray emission. The SN rate in starburst galaxies is high and $\gamma$-ray emission due to
relatively frequent SN explosions can be transient. The HE $\gamma$-ray luminosity of the starburst galaxy NGC 2403 above the
calorimetric bound was proposed to be caused by an additional component from SN 2004dj \citep[][]{ReferenceXiLiu}, which is one of the nearest and brightest SNe.
The search for VHE $\gamma$-ray emission from ten
core-collapse SNe observed with H.E.S.S. did not reveal significant emission \citep[][]{ReferenceSimoni}.
One of these SNe, SN 2011ja, occurred in NGC 4945. However,
the H.E.S.S. data set corresponding to the time interval within one year of SN 2011ja has a short exposure time of 3.4
hours. Another recent event, SN 2018ivc, occurred in NGC 1068 \citep[][]{Bostroem2020}. Given the distance to NGC 1068, the $\gamma$-ray flux from this SN is likely to be significantly smaller.

Apart from SNe, AGN in Seyfert galaxies can also produce variable $\gamma$-ray emission. For example, variable $
\gamma$-ray emission from the starburst galaxy NGC 3424 was attributed to an AGN \citep[][]{ReferencePeng}. For NGC 4945, the models have considered either the temporal or spectral characteristics. These models are discussed in the following.

\begin{figure}
\includegraphics[width=0.5\textwidth]{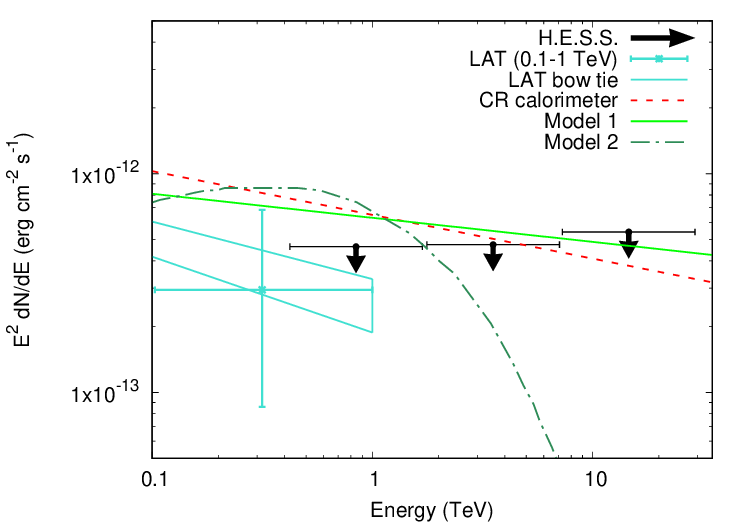}
\caption{Flux upper limits obtained from H.E.S.S. observations of NGC 4945. The calorimetric bound
        is from \citet[][]{ReferenceWang}; Models 1 and 2 are from \citet[][]{ReferenceWojaczynski} and \citet[][]{ReferenceJiang}, respectively.}
\label{plot_ngc4945}
\end{figure}

%\begin{itemize}

1. Temporal characteristics: \citet[][]{ReferenceWojaczynski} reported a hint
of an anti-correlation between the daily
hard X-ray and HE $\gamma$-ray fluxes from NGC 4945, suggesting a significant contribution from the AGN to the
HE $\gamma$-ray emission. They found that the lower the hard X-ray flux, the higher the HE $\gamma$-ray flux. This behavior is
reminiscent of the anti-correlation between hard X-ray and HE $\gamma$-ray fluxes in the X-ray binary Cygnus X-3 \citep[see,
e.g.,][]{ReferencePiano, ReferenceProkhorov}.
Additionally, they noted that the $\gamma$-ray spectrum hardens when the hard X-ray flux increases.
The difference between the photon indices in the two states is $\Delta\Gamma=0.36\pm0.11$. Thus, when extrapolated from
GeV to TeV energies, the $\gamma$-ray contribution from the subdominant component at 100 MeV becomes substantial at
TeV energies. Most of the H.E.S.S. observations of NGC 4945, corresponding to 80\% of the total exposure time, were performed 
when the source was in the high hard X-ray flux state. For this state, the flux extrapolated to VHE energies on the basis
of the spectral parameters taken from \citet[][]{ReferenceWojaczynski} is $F(>1\, \mathrm{TeV})=\left(4.3\pm0.8\right)
\times10^{-13}$~cm$^{-2}$s$^{-1}$. This extrapolated flux exceeds the flux upper limit, $F(>1\,
\mathrm{TeV})<3.1\times10^{-13}$~cm$^{-2}$s$^{-1}$, derived from the performed H.E.S.S. observations of the source in the
high hard X-ray flux state (see Fig. \ref{plot_ngc4945}). Model 1 shown in Fig. \ref{plot_ngc4945}
corresponds to the spectral parameters for the high hard X-ray flux state proposed by \citet[][]{ReferenceWojaczynski}.
This constraint suggests the presence of a cut-off in the $\gamma$-ray spectrum in the high hard X-ray state.
Verification of the two-state behavior using \textit{Fermi}-LAT and
\textit{Swift}-BAT data, in addition to those already used by \citet[][]{ReferenceWojaczynski}, is warranted.\\

2. Spectral characteristics: The two-state behavior can lead to a spectral hardening of the HE $\gamma$-ray spectrum when
analyzing the whole data set under the assumption
of a broken power-law spectrum. In this case, the lower-energy part is dominated by a soft $\gamma$-ray component
produced during the weak hard X-ray state, while the higher-energy part is dominated by a hard $\gamma$-ray component
produced during the high hard X-ray state. The broken power-law $\gamma$-ray spectrum can also result from the
presence of two different spectral components, as observed in the spectrum of another nearby galaxy, Centaurus A \citep[][]
{ReferenceSahakyan, ReferenceCenA}. In this light, the recent claim of a spectral hardening in NGC 4945 \citep[][]{ReferenceJiang},
based on an analysis of 11 years of the \textit{Fermi}-LAT data, is worth noting. Using a model with two hadronic
components, those authors performed a fit to the data and found that the higher-energy component has a very hard proton
power-law spectral index of $1.12^{+0.13}_{-0.19}$. This proton spectral index is much harder than the proton spectral index
in the nonlinear diffusive shock acceleration scenario, which is softer than 1.5 \citep[for a review, see][]{ReferenceMalkov}. The
differential flux upper limit at 0.84 TeV obtained from the H.E.S.S. observations of NGC 4945 is $E^2 \mathrm{d}N/\mathrm{d}
E<4.7\times10^{-13}$~erg cm$^{-2}$s$^{-1}$. Meanwhile, the flux value at this energy in the model with two
hadronic components from \citet[][]{ReferenceJiang} is $E^2 \mathrm{d}N/\mathrm{d}E=7.4^{+1.0}_{-3.4}\times10^{-13}$~erg cm$^{-2}$s$^{-1}$. Thus, the two-component model overpredicts the VHE $\gamma$-ray flux measured with H.E.S.S. (see, Model 2 in Fig. \ref{plot_ngc4945}). This implies that the values of the spectral parameters for the higher-energy component may differ from those derived by \citet[][]{ReferenceJiang}.

\subsection{VHE neutrinos from the surroundings of a SMBH}
\label{sect_neutrino}

Given that VHE neutrinos and accompanying $\gamma$ rays are the decay products of
pions created in the
$pp$ or $p\gamma$ interactions, the neutrino channel is used to place constraints on the VHE $\gamma$-ray production and
absorption.
Kinematics of charged and neutral pion decays establish the relation between the fluxes of neutrinos at energy
$E_{\nu}$ and $\gamma$ rays at energy $E_{\gamma}$, which is written as
\begin{equation*}
\frac{\mathrm{d}N_{i}}{\mathrm{d}E_i}=k_{i} \left(\frac{E_{i}}{\mathrm{TeV}}\right)^{-\Gamma}, \; \; i=\{\nu_{\mu}+\tilde{\nu}_{\mu},
\gamma\}
\end{equation*}
where $k_{\nu_{\mu}+\tilde{\nu}_{\mu}}\approx(0.71-0.16 \Gamma) k_{\gamma}$ for proton-proton ($pp$) interactions and $k_{\nu_{\mu}+
\tilde{\nu}_{\mu}}\approx2^{-\Gamma} k_{\gamma}$ for proton-photon ($p\gamma$) interactions (see \citealt[][]{ReferenceNeutrinoAstronomy,
ReferenceKappes, ReferenceAhlers}). Neutrinos are known to interact extremely weakly with matter and can leave
their production sites with essentially no attenuation. On the other hand, accompanying $\gamma$ rays can be heavily absorbed
within the source \citep[][]{ReferenceJelley, ReferenceBerezinsky}.

The search for point-like neutrino sources using ten years of IceCube data
resulted in an excess of 79$^{+22}_{-20}$ neutrinos
associated with NGC 1068 at a significance of 4.2$\sigma$ \citep[][]{IceCube3}. The reported best-fit flux averaged over the
data-taking period at a neutrino energy of 1 TeV is $\Phi^{1 \mathrm{TeV}}_{\nu_{\mu}+\tilde{\nu}_{\mu}}=(5.0\pm1.5_{\mathrm{stat}}
\pm0.6_{\mathrm{sys}})\times10^{-11}$~cm$^{-2}$ s$^{-1}$ TeV$^{-1}$ and the power-law index is $\Gamma=3.2^{+0.2}_{-0.2}
$. The neutrino flux from this source cannot be
explained by the starburst scenarios \citep[see][]{ReferenceEichmann}.
In the absence of $\gamma$-ray absorption, the expected $\gamma$-ray flux equivalent to the detected muon-neutrino flux would
be $\Phi^{1 \mathrm{TeV}}_{\mathrm{exp}, \gamma}\simeq2.2\times10^{-10}$~cm$^{-2}$ s$^{-1}$ TeV$^{-1}$ and
$4.4\times10^{-10}$~cm$^{-2}$ s$^{-1}$ TeV$^{-1}$ for $pp$ and $p\gamma$ interactions, respectively.
The flux upper limit set by the reported H.E.S.S. observations is significantly smaller. 
Below, Model A considers the entire energy range, but Model B restricts the energy range from 3 TeV to 30 TeV.
The latter energy range corresponds to that for IceCube observations scaled up by a factor of 2 (the average energy of $\gamma$ rays produced in $pp$ or $p\gamma$ interactions is twice larger than that of neutrinos).
For Model A, the H.E.S.S. upper limit translates to VHE $\gamma$-ray flux suppression factors of
$\gtrsim$3750 for $pp$ neutrino sources and $\gtrsim$7500 for $p\gamma$ neutrino sources. For the alternative model (Model B), the corresponding
VHE $\gamma$-ray flux suppression factors are $\gtrsim$150 and $\gtrsim$300, respectively.
Figure \ref{plot_ngc1068} shows the $\gamma$-ray fluxes corresponding to the IceCube neutrino flux from a $pp$ ($p\gamma$) neutrino source, reduced by a factor of 3750 (7500)
for Model A and 150 (300) for Model B due to absorption.
Due to the broader energy range, Model A is more constrained by the H.E.S.S. observations than Model B. For guidance purposes, the $\gamma$-ray spectra for the models A and B are shown in Fig. \ref{plot_ngc1068} as power laws corresponding to constant optical depths.

Additionally, the fact that the neutrino energy flux exceeds the HE $\gamma$-ray energy flux may require a significant attenuation of HE $\gamma$-rays from the neutrino production site. This implies the presence of a dense X-ray target photon field, which can exist only in the vicinity
of SMBHs. Due to the intense X-ray photon field, AGN coronae are one of the most plausible sites for neutrino
production \citep[][]{Inoue2020, Murase2020, ReferenceKheirandish, Eichmann2022, Mbarek2023, Neronov2023, Blanco2023, Inoue2024}. In the corona model, TeV $\gamma$ rays are cascaded down to sub-GeV energies \citep[e.g.,][]{Murase2024}. VHE neutrino production by protons accelerated in the inner regions of AGN-driven winds near the SMBH is possible for NGC 1068 \citep[][]{InoueCerruti}; see also \citet[][]{Peretti2023}.

By considering the connection between IceCube neutrinos and HE $\gamma$ rays in NGC 1068, \citet[][]{Murase2022} computed cascaded $\gamma$-ray spectra and showed that the neutrino production region likely lies at $R\lesssim 100~R_\mathrm{S}$. This is particularly the case in $p\gamma$ scenarios due to the Bethe-Heitler pair production process, where \citet[][]{Murase2022} obtained $R\lesssim 30~R_\mathrm{S}$, a scale which also corresponds to the estimated corona size in NGC 1068.

NGC 1068, along with TXS 0506+056, is visible near the horizon from the South Pole and IceCube has a good sensitivity for
detection of muon neutrinos from this direction \citep[][]{ReferenceIcecubeNGC1068excess}.
If detected in neutrinos \citep[e.g.,][]{ReferenceKheirandish}, the Circinus
galaxy will provide further insights into extreme astrophysical environments through comparing VHE neutrino and $\gamma$-ray fluxes.
The progress in enhancing IceCube's sensitivity for sources in the southern sky can expand searches for neutrino emission
to the Circinus galaxy \citep[][]{ReferenceMancina, Yu2023}.
The recently reported excess of neutrinos from the Circinus galaxy in the IceCube data highlights this progress \citep[][]{Yu2025}.
KM3NeT will reach higher sensitivity to muon-neutrino fluxes from the southern sky
and better angular resolution \citep[e.g.,][]{Ambrosone2024}.

\subsection{Application to CR energy losses and propagation}

\label{CalorimeterSection}

The calorimeter model requires that a substantial part of the energy injected in CR protons by SNe be lost to inelastic hadronic collisions
before CR escape from starburst galaxies \citep[][]{ReferenceThompson}.
Consequently, the star formation rate determines the CR yield and neutral pion decay dominates $\gamma$-ray production.
This model well describes $\gamma$-ray emission observed from
NGC 253 and M 82 \citep[e.g.,][]{ReferenceLacki, Abramowski2012}.

\subsubsection{Calorimetric efficiency}

\label{CalorimeterSection1}

It is important to clarify whether the calorimetric model can
describe the $\gamma$-ray production in NGC 1068, the Circinus galaxy, and NGC 4945. The $\gamma$-ray luminosities of two of these three galaxies, namely NGC 1068 and the Circinus
galaxy, in the GeV band were initially found to be significantly higher than those expected from the calorimeter model \citep[][]
{ReferenceLenain, ReferenceHayashida}. These high $\gamma$-ray luminosities were also treated as evidence that the
observed HE $\gamma$-ray emission from these two sources is dominated by the central AGN activity. For NGC 1068,
\citet[][]{ReferenceLenain} used the SN rate comparable to those of NGC 253 and M 82 (that is $\simeq$20$\pm10$ per century), 
but the $\gamma$-ray luminosity of NGC 1068 is higher by a factor of $\sim$10.
\citet[][]{ReferenceYoast} adopted similar SN rates in NGC 1068 and NGC 253 of 10 per century and 7 per century, respectively,
and also encountered difficulties in modeling the $\gamma$-ray flux of NGC 1068. These difficulties were resolved when the
different SN rates in NGC 1068 and NGC 253 of $35\pm9$
per century and $2.6\pm0.4$ per century, respectively, were applied by \citet[][]{ReferenceWang}. They found that the models both for NGC 1068 and NGC 4945 are
consistent with calorimetry, but they noted that the calorimetry relation fails for the Circinus galaxy.
In view of the $\gamma$-ray flux from the Circinus galaxy measured with \textit{Fermi}-LAT, as reported by
\citet[][]{ReferenceGuo, ReferenceEbrahim}, we re-examined the constraints set by \citet[][]{ReferenceWang}.
The re-examination shows that the calorimetric efficiency, which is a measure of the ratio of
the $\gamma$-ray energy output to the SN energy injected in CR protons,\footnote{The calorimetric $\gamma$-ray
luminosity limit here assumes an average CR acceleration energy per SN of $3\times10^{50}$ erg.} is within the interval of 41\% to 70\% at the 70\% confidence level. These values are similar to those obtained for NGC 1068 and NGC 4945.
The measured infrared luminosity of NGC 1068 implies its $\gamma$-ray luminosity -- via the $L_\mathrm{IR}$–$L_\mathrm{\gamma}$ correlation for star-forming galaxies \citep[][]{Ajello2020} -- that is significantly higher than those of NGC 253 and M82, consistent with the result from \citet[][]{ReferenceWang}.

The flux upper limits derived from the H.E.S.S. observations for
a proton spectral index of 2.2 set upper limits on the calorimetric efficiencies at $58\%$ and $70\%$ for NGC 1068 and NGC 4945, respectively. To set these limits the predictions from the model by \citet[][]{ReferenceWang} were used. These constraints are comparable to the confidence intervals for
calorimetric efficiencies of 41\% to 88\% for NGC 1068 and 40\% to 73\% for NGC 4945,
as derived by \citet[][see their Fig. 6]{ReferenceWang} from the $\gamma$-ray fluxes in the GeV band.
These calorimetric efficiencies are higher than those obtained for M82 (35\%) and NGC
253 (39\%).
This shows that the calorimeter model remains a viable explanation of $\gamma$-ray
emission from NGC 1068 and NGC 4945, considering the H.E.S.S. flux upper limits.
The calorimetric bounds (100\% efficiency) for NGC 1068 and NGC 4945 taken from \citet[][]{ReferenceWang}
are compared to the H.E.S.S. flux upper limit in Figs. \ref{plot_ngc1068} and \ref{plot_ngc4945}.

For Circinus, even with 100\% efficiency, the predicted fluxes are below the H.E.S.S. limits.
This is because both the infrared luminosity and the predicted VHE $\gamma$-ray flux for this galaxy are about half that of NGC 4945.

\subsubsection{Connection to UHECRs}

Although the UHECR sources have not yet been firmly identified, if the CR production rate in UHECR
sources extends as a power law down to energies of 1 TeV, then it can lead to VHE $\gamma$-ray emission via neutral pion decay.

Atomic nuclei make up $\sim$99\% of the CRs that are detected on Earth. The CR spectrum extends over a vast
range of energies from $10^{9}$ eV to $10^{21}$ eV.
The initial idea that Galactic CRs originate from SNRs was based on energy budget considerations \citep[][]
{BookGinzburg}. Proton acceleration to an energy above $10^{15}$ eV pushes shock acceleration theory to its limits when applied to SNRs \citep[][]{ReferenceBell}.
The Larmor radius of a proton with an energy of $10^{18}$ eV
gyrating in a $\mu$G magnetic field is about 1 kpc, so UHECRs are extragalactic in origin
since their Larmor radius is larger than the size of the Galaxy.

The correlation between starburst galaxies and UHECR arrival directions (4.2 $\sigma$)
may suggest the existence of sources accelerating CRs above $3.9\times10^{19}$ eV in these galaxies \citep[see, e.g.,][]{ReferenceAuger, ReferenceAuger2022}.
A model with 9.7\% of the UHECR flux above
$3.9\times10^{19}$ eV from nearby starburst galaxies (the remaining 90.3\% are isotropic) was found to be favored
\citep[][]{ReferenceAuger}. 
The starburst model can explain an UHECR hotspot in the direction of the Centaurus A/M83 group owing to the fact that NGC
4945 is a member of the Centaurus A subgroup.
The potential CR-UHECR sources
include starburst superwinds \citep[][]{ReferenceSuperwinds}, hypernova remnants \citep[][]{ReferenceHypernovae},
low-luminosity gamma-ray bursts \citep[][]{Zhang2018}, and kiloparsec-scale jets \citep[][]{ReferenceKpcjets}.

In the galaxy NGC 4945, CR protons with energies from tens of GeV to multi-TeV are trapped
by inelastic hadronic collisions and cannot escape. These CRs are scattered by small-scale magnetic 
irregularities and travel on a long, random-walk path through the dense galactic medium. 
Consequently, the combination of dense gas and slow CR diffusion is the primary factor 
trapping these lower-energy CRs. 
In contrast, UHECRs have larger gyroradii, making them less effectively
scattered by magnetic turbulence. They can navigate through the galaxy more directly 
(in a quasi-ballistic rather than diffusive regime) and escape \citep[e.g.,][]{Condorelli2023}. 
NGC 4945 acts as a CR calorimeter for the former, while it cannot confine the latter; 
this difference, therefore, establishes the relationship between the hadronic HE-VHE $\gamma$-ray emission
and the UHECR flux from this galaxy.

The proton luminosity of hypothetical CR-UHECR sources above $10$~GeV is estimated here for NGC 4945.
The CR spectral production rate in this source is assumed to follow a power-law
energy distribution with a spectral index $s$ harder than 2.7 below $3.9\times10^{19}$ eV.
The contribution from this galaxy to the total UHECR flux from starburst galaxies is adopted from \citet[][]{ReferenceAuger}.
This luminosity is then compared with
that provided from SNRs (for SN rates, see Sect. \ref{CalorimeterSection1}).
Since the extrapolated spectrum spans 10 
decades of energy, the CR luminosity of hypothetical CR-UHECR sources strongly depends on the spectral index.
The CR power produced by hypothetical CR-UHECR sources is comparable with the CR power produced by SNRs if the spectral index is 2.15 and more than 10 times the CR power produced by SNRs if the spectral index is 2.3 or softer.
Thus, the calorimetry limit from the H.E.S.S. observations of NGC 4945 (Sect. \ref{sect:obs})
is violated if CR-UHECR sources have soft spectra, $s>2.15$.
CR-UHECR sources with softer spectra would be too bright in the VHE $\gamma$-ray band, exceeding the H.E.S.S. constraints.

If UHECRs mostly consist of carbon, then the $\gamma$-ray fluxes induced by CR-UHECR sources may remain below the upper limits from H.E.S.S. observations when $s\simeq2.2$.

\section{Summary}
\label{sect:sum}

This paper describes VHE observations of the nearby composite Seyfert/starburst galaxies NGC 1068, the Circinus galaxy, and NGC 4945 with H.E.S.S. These galaxies have both an AGN and a starburst core.
Spiral galaxies hosting an AGN are potentially detectable $\gamma$-ray sources due to an
SMBH supplying the power that efficiently turns into radiation. Bursts of intense star formation in these three galaxies may result
in VHE $\gamma$-ray emission produced through the collisions of CR protons accelerated in core-collapse SNRs with interstellar gas.
Flux upper limits are obtained here from 168.5 hours, 18.4 hours, and 42.7 hours of H.E.S.S. observations toward NGC 1068, the Circinus galaxy, and
NGC 4945, respectively. These upper limits are among the most stringent constraints for these galaxies to date. The non-detection in VHE $\gamma$ rays has implications for the particle populations and
physical conditions in these systems. These upper limits constrain
(i) the number of energetic electrons inside the kiloparsec-scale bubbles in NGC 1068 and the Circinus
galaxy, (ii) the two-state behavior of NGC 4945, as previously suggested by \textit{Fermi}-LAT $\gamma$-ray
observations, (iii) the absorption of VHE $\gamma$-rays on soft photons in NGC 1068, in the context of neutrino detection,
(iv) the fraction of the SN explosion kinetic energy converted into the
energy of CRs in NGC 1068 and NGC 4945,
and (v) the spectral hardness of hypothetical CR-UHECR sources in NGC 4945, which produce both CRs interacting with gas and UHECRs observed at Earth. In conclusion, VHE $\gamma$-ray observations of composite Seyfert/starburst galaxies probe a broad range of energetic astrophysical phenomena, but higher sensitivity is required for detections.

\begin{acknowledgements} The support of the Namibian authorities
and of the University of Namibia in facilitating the construction and operation of
H.E.S.S. is gratefully acknowledged, as is the support by the German Ministry
for Education and Research (BMBF), the Max Planck Society, the German Research Foundation (DFG), the Helmholtz Association, the Alexander von Humboldt Foundation, the French Ministry of Higher Education, Research and Innovation, the Centre National de la Recherche Scientifique (CNRS/IN2P3 and CNRS/INSU), the Commissariat \`{a} l’\'{e}nergie atomique et aux \'{e}nergies alternatives (CEA), the U.K. Science and Technology Facilities Council (STFC), the
Irish Research Council (IRC) and the Science Foundation Ireland (SFI), the Knut
and Alice Wallenberg Foundation, the Polish Ministry of Education and Science,
agreement no. 2021/WK/06, the South African Department of Science and Technology and National Research Foundation, the University of Namibia, the National Commission on Research, Science \& Technology of Namibia (NCRST),
the Austrian Federal Ministry of Education, Science and Research and the Austrian Science Fund (FWF), the Australian Research Council (ARC), the Japan Society for the Promotion of Science, the University of Amsterdam and the Science Committee of Armenia grant 21AG-1C085. We appreciate the excellent
work of the technical support staff in Berlin, Zeuthen, Heidelberg, Palaiseau,
Paris, Saclay, T\"{u}bingen and in Namibia in the construction and operation of the
equipment. This work benefited from services provided by the H.E.S.S. Virtual Organisation, supported by the national resource providers of the EGI Federation.
\end{acknowledgements}

\bibliographystyle{aa}

\end{document}